\documentclass[conference]{IEEEtran}
\usepackage{color}
\usepackage{epsf}
\usepackage{times}
\usepackage{epsfig}
\usepackage{epstopdf}
\DeclareGraphicsExtensions{.png,.pdf}
\usepackage{graphicx}
\usepackage{algorithm}
\usepackage{algorithmic}
\usepackage{amsmath}
\usepackage{amssymb}
\usepackage{amsxtra}
\usepackage{multirow}
\usepackage{mathtools}
\usepackage{subcaption}
\usepackage{multirow}
\usepackage{comment}
\usepackage[normalem]{ulem}
\usepackage{tabu}
\usepackage[dvipsnames]{xcolor}
\usepackage{slashbox}
\usepackage{cite}
\usepackage{caption}
\usepackage[bottom]{footmisc}
\usepackage{pdfpages}

\begin{document}
\title{Scalable and Secure Architecture for Distributed IoT Systems}

\author{\IEEEauthorblockN{Najmeddine Dhieb$^1$, Hakim Ghazzai$^1$, Hichem Besbes$^2$, and Yehia Massoud$^1$}
\IEEEauthorblockA{\small $^1$School of Systems  \& Enterprises, Stevens Institute of Technology, Hoboken, NJ, USA \\
Email: \{ndhieb, hghazzai, ymassoud\}@stevens.edu}
$^2$University of Carthage, Higher School of Communications of Tunis, Tunisia\\
Email: hichem.besbes@supcom.tn
}

\maketitle
\thispagestyle{empty}

\begin{abstract}
Internet-of-things (IoT) is perpetually revolutionizing our daily life and rapidly transforming physical objects into an ubiquitous connected ecosystem. Due to their massive deployment and moderate security levels, those devices face a lot of security, management, and control challenges. Their classical centralized architecture is still cloaking vulnerabilities and anomalies that can be exploited by hackers for spying, eavesdropping, and taking control of the network. In this paper, we propose to improve the IoT architecture with additional security features using Artificial Intelligence (AI) and blockchain technology. We propose a novel architecture based on permissioned blockchain technology in order to build a scalable and decentralized end-to-end secure IoT system. Furthermore, we enhance the IoT system security with an AI-component at the gateway level to detect and classify suspected activities, malware, and cyber-attacks using machine learning techniques. Simulations and practical implementation show that the proposed architecture delivers high performance against cyber-attacks.

\end{abstract}\vspace{0.3cm}

\section{Introduction}
Internet-of-Things (IoT) technology has become highly auspicious to enhance automation, efficiency, and comfort level for users. Indeed, the number of IoT devices has widely increased. It is expected to exceed $8.4$ billion devices in 2020 and reach $20.4$ billion devices in 2022 engendering a tremendous amount of traffic and data sharing among data providers and consumers~\cite{8742551}. Although IoT technology proved its efficiency in different fields and became essential in many applications such as healthcare, remote monitoring, smart homes, and smart agriculture~\cite{7491206,8651860,8537381,8597264}, it remains under continuous upgrades and does not reach its maturity yet. The IoT environment is indeed still vulnerable and can be controlled by hackers who can use moderate security levels of hardware as well as firmware vulnerabilities to control those devices for espionage and eavesdropping.
\let\thefootnote\relax\footnotetext{\hrule \vspace{0.1cm} This  paper  is  accepted  for  publication  in  IEEE  Technology \& Engineering Management Conference (TEMSCON’20), Detroit, USA, jun. 2020.\newline
Personal  use  of  this  material  is  permitted.  Permission  from IEEE  must  be  obtained  for  all  other  uses,  in  any  current  or  future  media,including  reprinting/republishing  this  material  for  advertising  or  promotional purposes, creating new collective works, for resale or redistribution to servers or lists, or reuse of any copyrighted component of this work in other works.}
Often, data leakage takes place during transmission, share, or storage of data, which may entail serious problems for the IoT owners and users. Indeed, usually, IoT devices acquire basic security and authentication levels. From this perspective, some researchers suggested blockchain-based solutions so as to reinforce the authentication and identification processes~\cite{8355139} as well as data encryption  and sharing~\cite{8679354,8355227}. However, those suggestions are not immune against hackers cyber-attacks who execute malware and exploit device vulnerabilities to achieve their goals. 
Moreover, the connected devices can be used and monitored by hackers and cybercriminals to create sophisticated cyber-attacks which may lead to dangerous and fatal consequences. Recently, Mirai malware used IoT devices and targeted Domain Name Service (DNS) servers in order to generate a sophisticated Distributed Denial of Service (DDoS) attacks which affected the internet service for a large number of users such as Netflix, GitHub, and Reddit and causes huge losses for those companies~\cite{7971869,8813605}. 

One of the most effective means to secure IoT devices and services is providing an end-to-end secure IoT architecture and uncircumventable access control for IoT devices. Artificial Intelligence (AI) has been widely used in many industrial domains for its efficiency in upgrading IoT devices with sophisticated smart features~\cite{8691103}. In this context, we propose a novel architecture for IoT devices combining  blockchain and AI technologies not only in order to decentralize data storage and protect shared data into the IoT network but also to enhance its performance and efficiency against malware and cyber-attacks. This work investigates the power of machine learning techniques as well as the efficiency of blockchain technology in order to improve privacy, data sharing, and security for IoT devices and smart city infrastructure which are vulnerable and can be used for sinful activities. We advocate the use of the permissioned blockchain to share and store the IoT devices data for its compatibility in IoT distributed architecture, where unlike some suggested approaches~\cite{8355227,7890132}, the connected devices do not participate in the mining process and decision-making due to their limited computation power. Then, we apply machine and deep learning algorithms such as Artificial Neural Networks (ANN), XGBoost, decision tree, and naive bayes for malware detection and classification, running on nodes participating in the blockchain network to control and  detect suspicious activities. Our simulation results illustrate the efficiency of the proposed architecture for IoT privacy, data sharing, and malware detection into the proposed blockchain network.

\begin{figure*}[t]
\vspace{0.3cm}
\centering

     \includegraphics[scale=0.4]{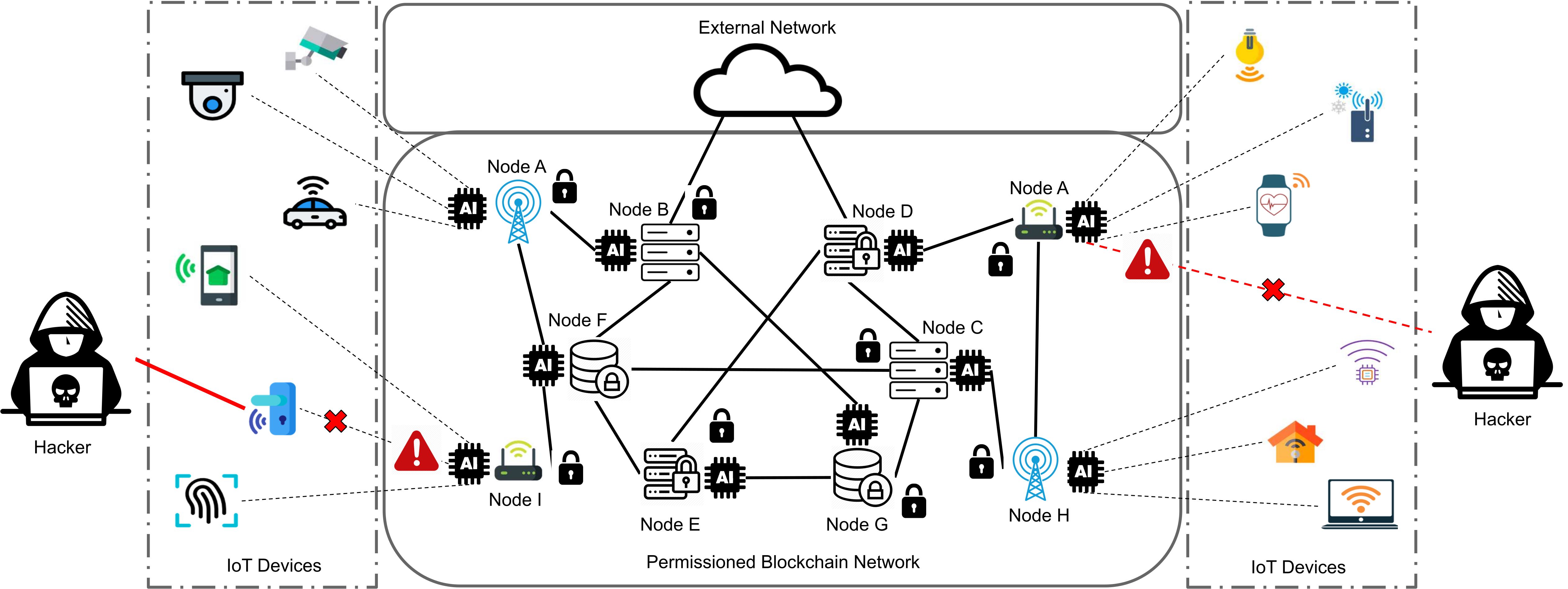}
    \caption{Artificial intelligence and blockchain based architecture for heterogeneous IoT systems. The solid lines refer to connection and communication links between nodes participating into the blockchain. The dashed line reflects the communication links between IoT devices and edge nodes. Red lines (solid and dashed) represent vulnerable connections and suspected IoT communications.}\label{blockchainiotarch}
\end{figure*}

\section{Related work}\label{sec3}
In this section, we discuss some of the state-of-art techniques and technologies applied to IoT systems while highlighting the novelty of this work with respect to existing studies.

A system integrating blockchain, Service-oriented Architecture (SoA), and Key Performance Indicator (KPI) was suggested by W. Viriyasitavat et al. to ensure data validity and deal with the heterogeneity, uncertainties, and mobility of devices~\cite{8764459}. Blockchain was also applied in~\cite{8843900} for data sharing into multiple distributed parties in industrial IoT. Federated learning was integrated in consensus process of the blockchain to build data models and share them into the blockchain network. 
Liu et al. developed a shared data architecture based on blockchain technology and deep reinforcement learning for Industrial IoT devices~\cite{8594641}.
A decentralized architecture based on permissioned blockchain was suggested in~\cite{8824921} in order to share and secure the circulating data in smart city infrastructure. Hyperledger Sawtooth was utilized to overcome the infrastructural challenges for the smart city deployment. In addition, a supplemental module was developed to automate and reduce the blockchain deployment processes. The previous solutions may provide a generic architecture for IoT services and integrate different cutting-edge technologies. However, they are unable to protect the IoT environment and the blockchain network from hackers who may control the network by bypassing the moderate hardware security and exploit vulnerabilities in IoT firmware to achieve their atrocious activities.

In order to emulate IoT services, a machine learning approach was suggested by M. Pahl et al. based on inter-service communication observations~\cite{8584985}. They developed an inter-service communication model for micro-services analysis between different IoT devices. In addition, they suggested a continuous correlation algorithm for different observations within the IoT devices.
Trust list model was suggested by K. Kataoka et al. to automate the verification and trusting of IoT devices and services~\cite{8355139}. In this context, a trust list presenting a distributed trust between the different IoT-related stakeholders is developed. In this architecture, blockchain and software-defined network (SDN) were integrated to automate and enforce the IoT authentication processes as well as circulating the trusted IoT services and devices along the different stakeholders presented in the network. 
An IoT malware detector for large-scale networks was proposed in~\cite{8767194}. A machine learning algorithm is executed on the user access gateway in order to detect the suspicious activities based on patterns scanning traffic. Additionally, a policy module was developed to determine the needed actions for malicious packets. Moreover, a database was applied to store the scanned patterns in order to update or receive them when needed. A malware detection architecture was developed in~\cite{8721053} for android IoT devices. In this architecture, machine learning techniques were applied to extract malware information received from an android device and store them into a blockchain network. The blockchain stores the malware features and share them into the distributed ledger. M. Shen et al. suggested Paillier public-key crypto system in order to encrypt and protect the data provided from IoT devices~\cite{8653362}. In addition, they implemented the support vector machine (SVM) algorithm to train the machine learning model directly from the blockchain network. Data provided from the IoT device is encrypted then saved and distributed on different distributed ledgers.

Although the previous studies suggested some malware detection approaches based on intelligent systems and machine learning techniques, they were unable to guarantee the authenticity and the integrity of data transmitted from the IoT devices into the IoT network. In fact, the architecture of typical IoT environments relies on brokered communication models, which have limited scalability and are exposed to multiple vulnerabilities due to the centralized authority for identification, authentication, and storage.

\section{Proposed Architecture}\label{sec3'}

The proposed architecture is essentially built on the idea of exploiting AI and blockchain advantages to design a more robust and resilient system ensuring high-level of security and scalability.

Typical public blockchain solution presents multiple drawbacks which keep it far away of being used for generic IoT platforms due to the limited resources and computation power of IoT devices. In fact, this limitation prevents those devices from being effective miners in the blockchain network. Also, due to the massive deployment of those devices and their basic security levels, the blockchain network can be manipulated if more than half of these nodes were accessed by unauthorized entities. For those reasons, we propose to employ a novel permissioned  blockchain network architecture combined with AI technologies as illustrated in Fig.~\ref{blockchainiotarch}. 

In traditional IoT environment, hackers may exploit firmware and hardware vulnerabilities present in gateways, servers, and IoT devices such that they can manipulate them and use their amassed power to achieve different cyber-attacks such as DDos which may lead to temporary denial of service in the network. However, in the proposed architecture, AI modules integrated in the IoT environment can detect suspicious and abnormal activities  occurring in the network as well as protect those devices from different cyber-attacks. IoT devices submit their data to gateways which are also considered as nodes participating in the blockchain network with other servers and nodes sharing the same IoT network. Submitted information are verified, validated, and integrated into the network by authorized nodes. Also, the proposed solution ensures a blockchain based architecture for all devices and nodes participating into an IoT environment as well as the ability to extend and connect this network to other blockchain or any external network. The proposed architecture ensures decentralization, high scalability, and good performance thanks to the permissioned blockchain features as well as sufficient security levels, malware, and cyber-attacks detection ensured by the efficient AI modules implemented into all blockchain nodes. The malware detection module is performed on IoT gateways and nodes participating into the blockchain network. This assumption ensures early and real-time detection as well as reduces the risk to any cyber-attack attempt for the blockchain network or any malware execution from the IoT devices or between the network nodes.  
\begin{figure*}[t]
\vspace{0.3cm}
\centering
      \includegraphics[scale=0.08]{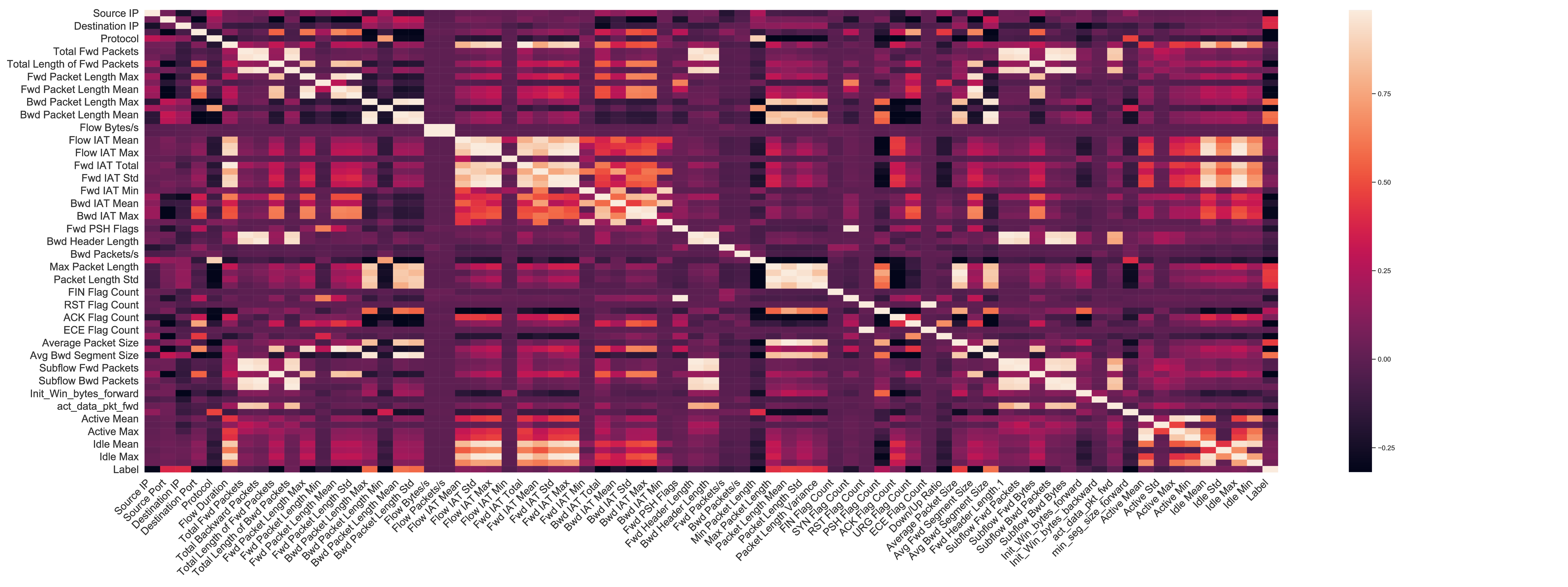}
    \caption{Correlation heat-map matrix for the selected features.} 
     \label{FeaturesCorrelation2}\vspace{-0.cm}
\end{figure*}

\section{Permissioned Blockchain for IoT Network}
A blockchain is a growing, distributed, and immutable time-stamped  data records that are linked using cryptography, managed, and distributed over different nodes participating in that network. Originally, blockchain was designed to record transactions for cryptocurrency systems. However, the use of this technology has been widely extended and used in many other fields such as industry and healthcare~\cite{8813031,9046765}. In literature, there are two major types of blockchain architecture: permissioned and permission-less blockchain networks, based on the application scenario, transactions types, and access rules. Also, different blockchain platforms were developed such as Ethereum, Ripple, and Hyperledger to emulate the blockchain architecture, mining process, and the access restrictions for users~\cite{8813712}.

In permission-less blockchain, \textit{aka} public blockchain, anyone can join the blockchain network, submit transactions, participate in the mining process, and leave the network without any restriction or permission needed in advance. This type of architecture ensures an open and decentralized data sharing network between all nodes participating into the blockchain. Moreover, all submitted transactions should be verified and validated before creating the block and being integrated into the blockchain. This process is ensured by a consensus protocols. Usually, in public blockchain, the Proof-Of-Work (PoW) consensus mechanism is used to validate transactions and create new blocks to the chain.

For permissioned blockchains, sets of rules are defined and shared between all nodes participating into the blockchain network in order to control access and manage this network. In fact, this type of blockchains is considered more secure and restricted compared to the public one. Commands and management rules are not authorized for all nodes participating into the network, but they are assigned to certain set of trusted nodes which have these authorities. Miners are also authorized nodes in the network and they are the only entities that have the authority to create, validate, and integrate new blocks into the blockchain. Practical Byzantine Fault Tolerance (PBFT)~\cite{castro1999practical} is the commune consensus protocol for this type of blockchain. The PBFT algorithm is an improvement of the BFT, which has the ability to be run in asynchronous environment and also ensures high-speed transaction processing while solving the Byzantine general problems.

For privacy and efficiency concerns, we advocate the use of a permissioned blockchain, more precisely, the hyperledger fabric which is an implementation of distributed ledger technology that ensures high level of security, scalability, and performance~\cite{androulaki2018hyperledger}. This type of permissioned system ensures strong identity management between different IoT nodes, distinguishes between users and validators, and affects each user its exact role into the network. In fact, hyperledger fabric provides a membership identity service that controls member IDs and authenticates users participating into the blockchain network. Access and permission rules provide an additional security layer for the proposed IoT network, where these devices can participate into the blockchain network with restricted authorities and control to minimize data leakage from the IoT devices. In addition, thanks to its modular design, this system presents better performance and scalability, where data submission requires less validation time while maintaining decentralized storage and better security and privacy levels. Nevertheless, this architecture is not sufficient to protect the network from malware and cyber-attacks especially the ones initiated through IoT devices. Therefore, we add another layer of security based on AI and machine learning to cope with those problems unsolved by permissioned blockchain.


\section{Machine Learning for Malware Detection and Classification}
In this section, we investigate the application of AI and machine learning algorithms into the blockchain network in order to enhance system security and its effectiveness against malware and cyber-attacks. The proposed approach relies on implementing efficient and sophisticated AI modules to automatically detect specious activities and protect all nodes participating into the blockchain network. 

We introduce the proposed methodology for data cleaning and mining as well as the feature engineering and model building processes so as to build an efficient and robust AI module. This solution cannot only detect malware and cyber-attack but also classify them into different categories in real-time and can prevent hackers from reaching their sinful goals.

Usually, numerical datasets have corrupted information or missing values which may affect the quality as well as the machine learning model accuracy. In this case, data cleaning and processing techniques are applied for the dataset in order to deal with this problem which can be caused during data insertion, transport, or storage of information into the database. During this process, we identify incomplete submissions, incorrect values, and missing features. Then, we correct, replace, or remove inaccurate records from the dataset so as to reduce training time, enhance data quality, and improve the machine learning model efficiency.

After cleaning the data, we explore and analyze its features. In fact, in order to get robust and accurate detection model, we have to train and validate the machine learning algorithm with relevant features from the dataset. In this context, we  analyze the dataset, summarize its information, and extract the most important features using visualization and statistics methods such as feature correlation. Fig.~\ref{FeaturesCorrelation2} presents the features correlation matrix used in this work to analyze the dataset features. The higher the correlation between the different features is the lower the information that we can get from these features are. As a result, we eliminate redundant and weak features such as 'Flow ID', 'PSH Flags', 'URG Flags', and 'Bulk Rate'.

Before we build the detection model, we prepare the cleaned data for machine learning. Most of the machine learning algorithms do not support different types of data such as text and structural entities during training phase, as a result we carefully prepare and adapt those features so that we transform them into other formats while preventing the creation of any additional numerical relationship between different values. 

We convert the malware and cyber-attacks detection into a machine learning classification problem. With this assumption, we categorize different types of malware and attack into different classes. To this end, we adopt a multi-class categorical classification technique and investigate relevant machine learning algorithms to solve this problem. 
\begin{figure*}[t]
\vspace{0.2cm}
\centering
\begin{minipage}{.5\textwidth}
  \centering
  \includegraphics[width=\linewidth]{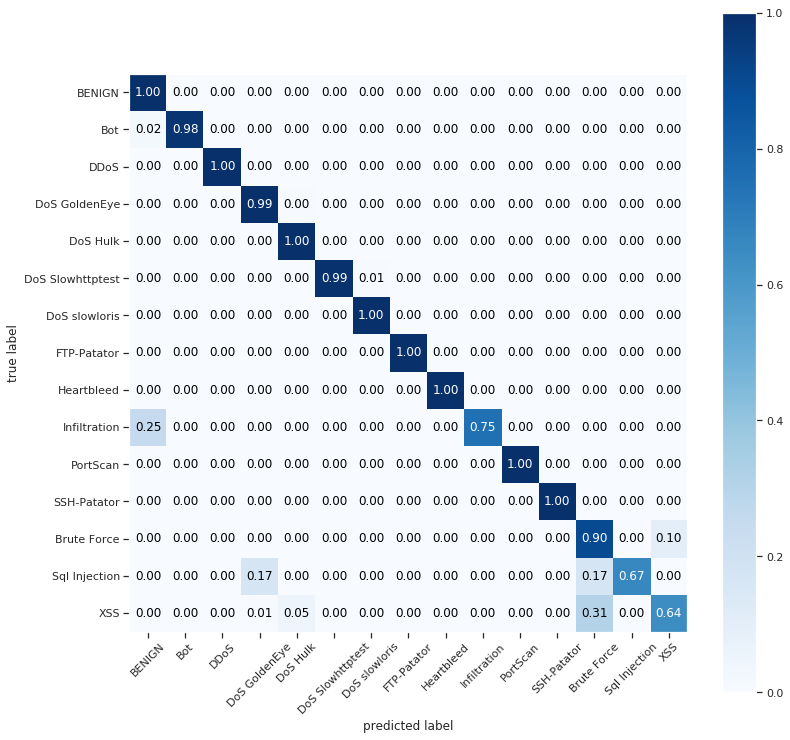}
  \captionof{figure}{XGBoost normalized confusion matrix.}
  \label{NormalizedCM1}
\end{minipage}%
\begin{minipage}{.5\textwidth}
  \centering
  \includegraphics[width=\linewidth]{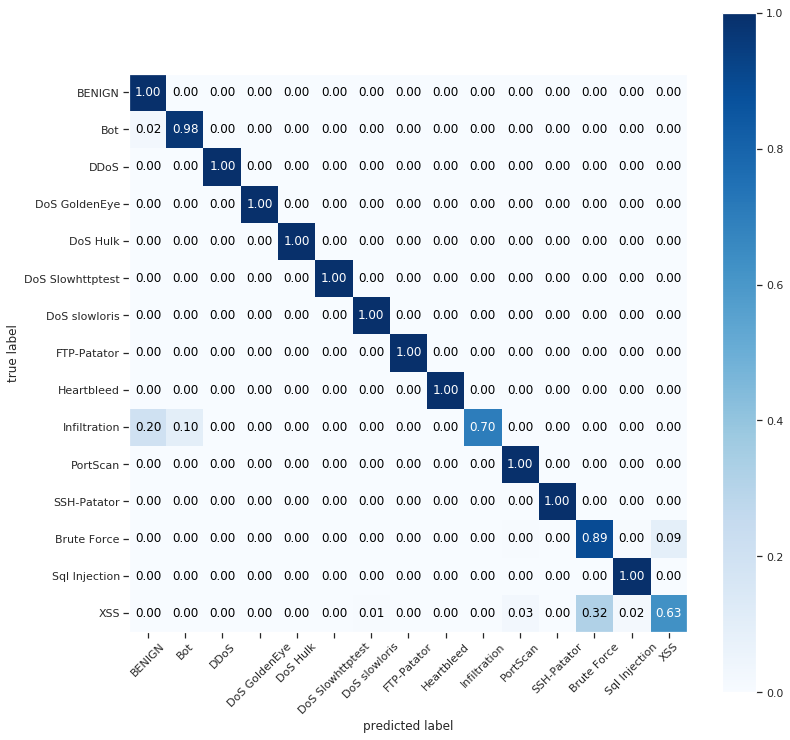}
  \captionof{figure}{Very fast decision tree normalized confusion matrix.}
  \label{NormalizedCM2}
\end{minipage}
\end{figure*}

\section{Performance Evaluation and Implementation}\label{sec5}
In this section, we simulate, implement, and evaluate the performance of the proposed IoT architecture based on blockchain technology and AI. For the malware detection, we use a real dataset from the Canadian Institute of Cybersecurity~\cite{dataset}. The datasets contain more than three million samples of different data traffic cyber-attacks during five days. After merging, cleaning, and processing the data, we obtain a dataset with more than two million and 8 hundred thousand instances spanning fifteen different cyber-attack classes. Table~\ref{tab1} describes the data and different attacks used to train the machine learning model. We define by \textit{benign} the normal traffic coming from the IoT devices and \textit{malicious} all suspected traffic that includes malware or unusual behavior from the devices connected to that node. The proposed model can recognize normal and suspected traffic as well as detect and distinguish fourteen cyber-attack categories. 

\begin{table}[t]
\begin{center}
\caption{\label{tab1} Dataset overview}
\addtolength{\tabcolsep}{-4pt}
\resizebox{8.8cm}{!}{
\begin{tabular}{cccclll}
\cline{1-4}
\multicolumn{1}{|c|}{\textbf{Traffic category}}              & \multicolumn{1}{c|}{\textbf{Attack type}}      & \multicolumn{1}{c|}{\textbf{Samples number}} & \multicolumn{1}{c|}{\textbf{Total}}                   &  &  &  \\ \cline{1-4}
\multicolumn{1}{|c|}{\textbf{Normal traffic}}                & \multicolumn{1}{c|}{Benign}           & \multicolumn{1}{c|}{2272688}        & \multicolumn{1}{c|}{2272688}                 &  &  &  \\ \cline{1-4}
\multicolumn{1}{|c|}{\textbf{Botnet}}                       & \multicolumn{1}{c|}{Bot}              & \multicolumn{1}{c|}{1966}           & \multicolumn{1}{c|}{1966}                    &  &  &  \\ \cline{1-4}
\multicolumn{1}{|c|}{\multirow{2}{*}{\textbf{Brute force}}} & \multicolumn{1}{c|}{FTP Patator}      & \multicolumn{1}{c|}{7938}           & \multicolumn{1}{c|}{\multirow{2}{*}{13835}}  &  &  &  \\ \cline{2-3}
\multicolumn{1}{|c|}{}                             & \multicolumn{1}{c|}{SSH Patator}      & \multicolumn{1}{c|}{5897}           & \multicolumn{1}{c|}{}                        &  &  &  \\ \cline{1-4}
\multicolumn{1}{|c|}{\multirow{6}{*}{\textbf{Dos/DDos}}}    & \multicolumn{1}{c|}{DDos}             & \multicolumn{1}{c|}{128027}         & \multicolumn{1}{c|}{\multirow{6}{*}{379750}} &  &  &  \\ \cline{2-3}
\multicolumn{1}{|c|}{}                             & \multicolumn{1}{c|}{Dos GoldenEye}    & \multicolumn{1}{c|}{10293}          & \multicolumn{1}{c|}{}                        &  &  &  \\ \cline{2-3}
\multicolumn{1}{|c|}{}                             & \multicolumn{1}{c|}{Dos Hulk}         & \multicolumn{1}{c|}{230124}         & \multicolumn{1}{c|}{}                        &  &  &  \\ \cline{2-3}
\multicolumn{1}{|c|}{}                             & \multicolumn{1}{c|}{Dos slowhttptest} & \multicolumn{1}{c|}{5499}           & \multicolumn{1}{c|}{}                        &  &  &  \\ \cline{2-3}
\multicolumn{1}{|c|}{}                             & \multicolumn{1}{c|}{Dos slowloris}    & \multicolumn{1}{c|}{5796}           & \multicolumn{1}{c|}{}                        &  &  &  \\ \cline{2-3}
\multicolumn{1}{|c|}{}                             & \multicolumn{1}{c|}{Heartbleed}       & \multicolumn{1}{c|}{11}             & \multicolumn{1}{c|}{}                        &  &  &  \\ \cline{1-4}
\multicolumn{1}{|c|}{\textbf{Infiltration}}                 & \multicolumn{1}{c|}{Infiltration}     & \multicolumn{1}{c|}{36}             & \multicolumn{1}{c|}{36}                      &  &  &  \\ \cline{1-4}
\multicolumn{1}{|c|}{\textbf{PortScan}}                     & \multicolumn{1}{c|}{PortScan}         & \multicolumn{1}{c|}{158930}         & \multicolumn{1}{c|}{158930}                  &  &  &  \\ \cline{1-4}
\multicolumn{1}{|c|}{\multirow{3}{*}{\textbf{Web attack}}}  & \multicolumn{1}{c|}{Brute Force}      & \multicolumn{1}{c|}{1507}           & \multicolumn{1}{c|}{\multirow{3}{*}{2180}}   &  &  &  \\ \cline{2-3}
\multicolumn{1}{|c|}{}                             & \multicolumn{1}{c|}{SQL Injection}    & \multicolumn{1}{c|}{21}             & \multicolumn{1}{c|}{}                        &  &  &  \\ \cline{2-3}
\multicolumn{1}{|c|}{}                             & \multicolumn{1}{c|}{XSS}              & \multicolumn{1}{c|}{652}            & \multicolumn{1}{c|}{}                        &  &  &  \\ \cline{1-4}
\multicolumn{1}{l}{}                               & \multicolumn{1}{l}{}                  & \multicolumn{1}{l}{}                & \multicolumn{1}{l}{}                         &  &  & 
\end{tabular}
 }
\end{center}\vspace{-0.6cm}
\end{table}
To train the AI module, we use different classification machine learning algorithms such as XGBoost, Decision tree, and Gaussian Naive Bayes. Also, we implement a deep learning-based solution to detect malware and suspected traffic. In order to evaluate the performance of these algorithms and methodologies, we measure the accuracy, precision, recall, and F1-score as shown in Table~\ref{tab2}. Accuracy shows a high performance for the decision tree algorithm. XGBoost algorithm achieves a close accuracy compared to the decision tree. However, it shows better performance for the other evaluation metrics.
The notable performance of the decision tree and XGBoost algorithms are validated by the confusion matrices given in Fig.~\ref{NormalizedCM1} and Fig.~\ref{NormalizedCM2} which highlight the malware detection efficiency of those algorithms for all cyber-attack types existing in the testing dataset.

\begin{table}[t]
\begin{center}
\caption{\label{tab2} Performance of the malware detection module (macro average) }\vspace{-0.2cm}
\addtolength{\tabcolsep}{-4pt}
\resizebox{8.8cm}{!}{
\begin{tabular}{|c||c|c|c|c|c|}%
  \hline
 \textbf{Classifier} & \textbf{Accuracy (\%)} & \textbf{Precision} & \textbf{Recall} & \textbf{F1-Score}  \\
  \hline
  \textbf{ANN}& 98.84 & 0.96 & 0.85 & 0.90\\
  \hline
   \textbf{Decision Tree}& {99.97} & 0.92 & 0.95 & 0.92 \\
  \hline
       \textbf{Naive Bayes}& {37.68} & {0.50} & {0.57} & {0.39} \\
  \hline
     \textbf{XGBoost}& {99.96} & {0.96} & {0.93} & {0.94} \\
  \hline
\end{tabular}
 }
\end{center}\vspace{-0.3cm}
\end{table}

\begin{figure*}[t]
\vspace{0.3cm}
\centering
\begin{minipage}{.5\textwidth}
  \centering
  \includegraphics[width=8.5cm]{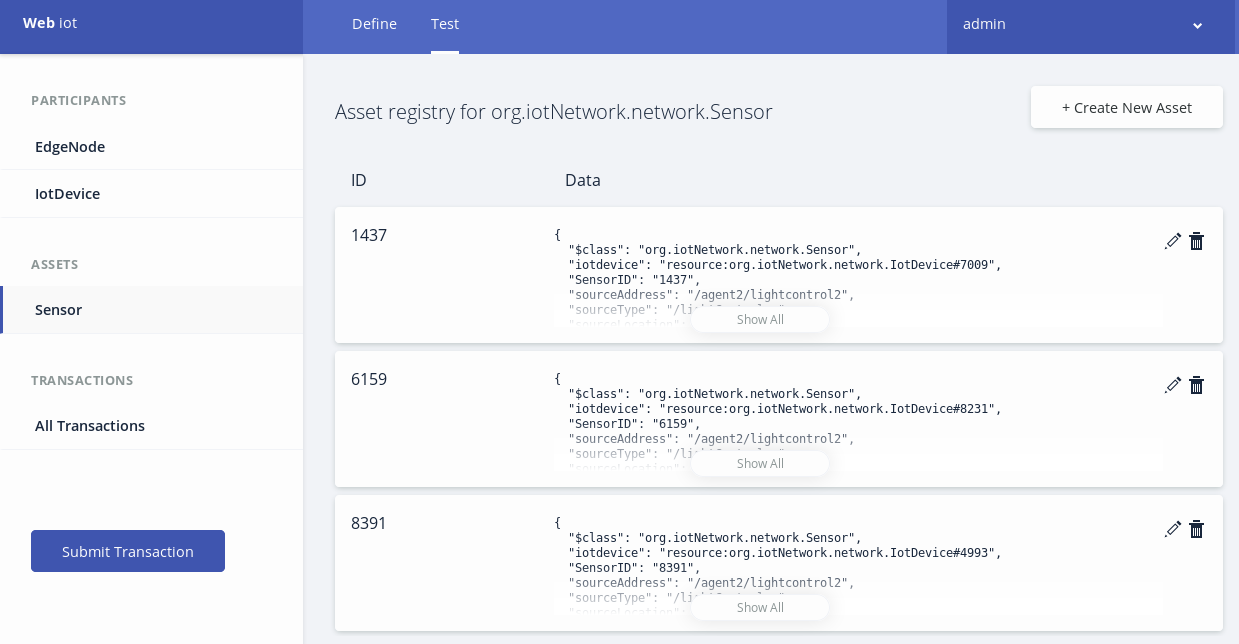}
  \captionof{figure}{Edge node interface.}
  \label{block1}
\end{minipage}%
\begin{minipage}{.5\textwidth}
  \centering
  \includegraphics[width=8.5cm]{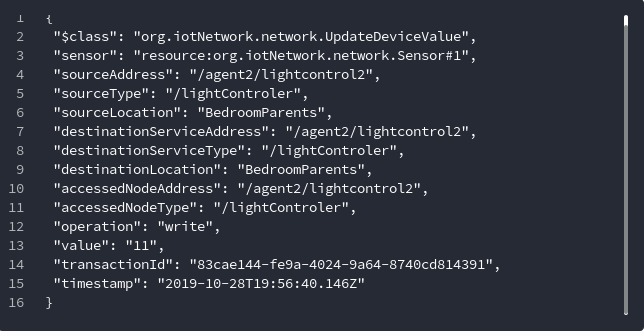}
  \captionof{figure}{IoT device transaction.}
  \label{block2}
\end{minipage}\vspace{-0.0cm}
\end{figure*}

For the blockchain architecture, we present and simulate the permissioned blockchain network using hyperledger fabric~\cite{8069090} where roles, authorities, and identities of members are known to all participants. In this network, we take all gateways, servers, and nodes participating into the IoT environment as participants. Also, we define the IoT devices as assets participating into the network. Moreover, we assume that each IoT device sends the information to the gateway by submitting a transaction to the edge node which receives data from this device then verifies and stores it into the blockchain network. 

All the blockchain simulations and applications are developed using hyperledger composer playground module of the hyperledger fabric framework. Fig.~\ref{block1} presents the graphical application interface for the blockchain network, where IoT devices, that might enclose multiple sensors, submit transactions to update their reported measures. As illustrated in the figure, each sensor has its own ID and is associated to one IoT device, which is also identified by its unique ID. For example, the sensor $1437$ is attached to the IoT device $7609$ and its last value equal to $23$. After each measure taken by the sensor, the IoT device submits a transaction to the edge node in order to update the sensor values and save them into the blockchain. In Fig.~\ref{block2}, we illustrate an example of a transaction made by the sensor with ID $1$ to update its value to $11$ into the blockchain.

\section{Conclusion}\label{sec6}
In this study, we introduced a novel architecture for IoT networks based on blockchain and AI technologies to decentralize authorities, enhance data sharing, and protect them from cyber-attacks. Due its compatibility with the distributed nature of IoT systems, we employed permissioned blockchain to share and store the IoT devices data. Then, we supported edge and back-end entities with AI modules where machine and deep learning algorithms such as ANN and decision tree-based algorithms are implemented to detect and classify suspicious activities in the blockchain network. Through implementation and simulations, we evaluate the efficiency of the proposed architecture in detecting and classifying cyber-attacks using practical and real-world datasets where decision tree based models show better performance compared to other state-of-the-art algorithms. 
As a future work, we aim to strengthen the proposed architecture by other AI modules capable of not only detecting and classifying cyber-attacks but also automatically recovering and helping make cyber-attack response decisions.

\bibliographystyle{IEEEtran}
\bibliography{references}
\end{document}